\documentclass{jps-cp}
\usepackage{txfonts} %Please comment out this line unless the txfonts package is availabe in your LaTeX system.
\usepackage{graphicx}
\title{The Upgrade of the CMS Tracker at HL-LHC}

\author{Alessandro \textsc{La Rosa}$^{1}$ on behalf of CMS Tracker collaboration}

\inst{$^{1}$CERN, Geneva 23, CH-1211 Switzerland}

\email{alessandro.larosa@cern.ch}

\abst{In the high luminosity scenario of the LHC (HL-LHC), which will bring the instantaneous luminosity up to 7.5\,$\times$\,$10^{34}$\,cm$^{-2}$s$^{-1}$,  ATLAS and CMS will need to operate at up to 200 interactions per 25\,ns beam crossing and reaching up to 4000\,fb$^{-1}$ of integrated luminosity. To achieve their physics goals the experiments will need to improve the tracking and vertexing capability and the ability to selectively trigger on specific physics events at reasonable thresholds. The upgrade of the CMS Tracker requires designing new inner and outer tracking detectors to cope with the increased luminosity and to implement first trigger level functionality. This paper describes the new layout and the technological choices together with some highlights of research and development activities.}

\kword{HL-LHC tracking detectors, silicon pixel detectors, silicon strip detectors}

\begin{document}
\maketitle

\section{Introduction}
The High-Luminosity Large Hadron Collider (HL-LHC~\cite{HL-LHC}) at CERN is expected to collide protons at a centre-of-mass energy of 14 TeV and to reach the unprecedented peak instantaneous luminosity of 7.5\,x\,$10^{34}$\,cm$^{-2}$s$^{-1}$ with an average number of pileup interactions up to 200. This will allow the ATLAS~\cite{ATLAS} and CMS~\cite{CMS} experiments to collect integrated luminosities of 3000\,fb$^{-1}$  during the project baseline lifetime, with the possibility to reach 4000\,fb$^{-1}$ as ultimate scenario. To cope with this very challenging scenario the CMS detector will be substantially upgraded~\cite{TKTDR} before the start of the HL-LHC. The CMS tracking detector will have to be replaced in order to fully exploit the delivered luminosity and cope with the demanding operating conditions. The new detector will provide robust tracking as well as input for the first level trigger (L1).

\begin{figure}[!ht]
\centering
\includegraphics[scale=0.3]{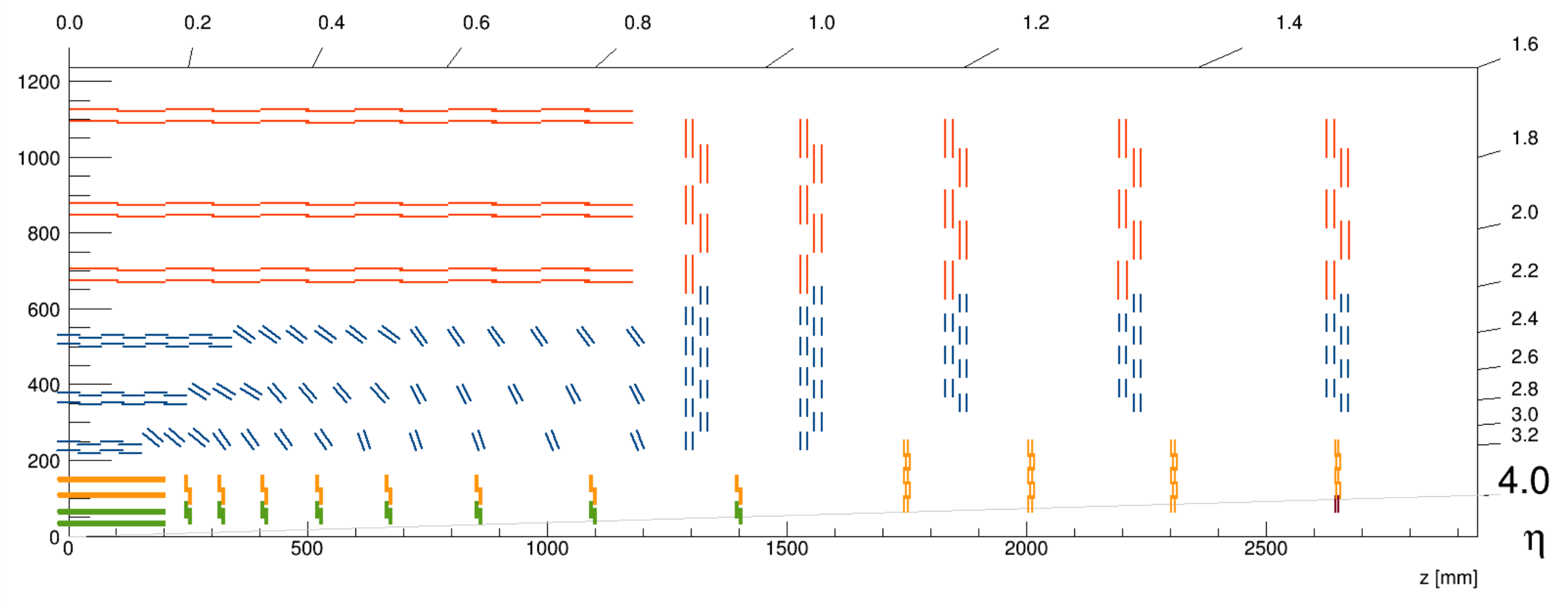}
\caption{Sketch of one quarter of the CMS Phase-2 Tracker layout in r-z view. In the Inner Tracker the green lines correspond to pixel modules with two readout chips (ROCs) and the yellow  and brown lines to pixel modules with four ROCs. In the Outer Tracker the blue and red lines represent the two types of modules,  equipped with macro-pixel-strip and strip-strip sensors, respectively.}
\label{fig:TK-layout}
\end{figure}

The CMS Phase-2 tracker  will consist of about 200\,m$^{2}$ of silicon modules and will be composed of two sub-detectors:  the Inner Tracker (IT)  made of silicon pixel modules and  the Outer Tracker (OT) made of a combination of silicon modules  with strip and macro-pixel sensors.  A longitudinal view of one quarter of the new detector layout is shown in Fig.\,\ref{fig:TK-layout}.

The main requirements for the new tracker can be summarised as follows: radiation tolerance to be fully efficient up to the expected integrated luminosity; increased granularity to ensure excellent tracking performance in the presence of a high level of pileup (occupancy less than 1\% for the OT and 0.1\% for the IT); reduced material in the tracking volume; contribution of tracking information to the L1 trigger (only for OT);  large readout bandwidth and deep front-end buffers for higher rate (750 kHz) and longer latency (12.5\,$\mu$s) of the L1 trigger system; extended tracking acceptance up to $\mid$$\eta$$\mid$ $\sim$4 for efficient pile-up mitigation and better track reconstruction in the forward region.

%

%%%%%%%%%%%%%%%%%%%%%

\section{Inner Tracker system}

The Inner Tracker consists of four barrel layers (TBPX) plus eight small disks (TFPX) and four large disks (TEPX) per side, covering a surface of 5\,m$^{\mathrm{2}}$  with 3900 hybrid pixel modules featuring about 2 billion readout channels. In addition to extending the tracking acceptance up to $\mid$$\eta$$\mid$ $\sim$4, the TEPX disks  are employed to perform real time luminosity measurements. The innermost rings of the outermost TEPX disks are exclusively used for beam background and luminosity measurements with independent readout and control systems. 
The expected hadron fluences and total ionising dose for the innermost pixel layer are 3.5\,$\times$\,$10^{16}$\,n$_{\mathrm{eq}}$\,cm$^{-2}$ and 1.9\,Grad, respectively. A replacement of  the innermost pixel layer and the innermost ring of the TFPX  will be required during the long shutdown (LS) number 5 of the current HL-LHC schedule, irrespective of which sensor technology will be selected. 

\begin{figure}[h!]
\centering
\includegraphics[scale=0.45]{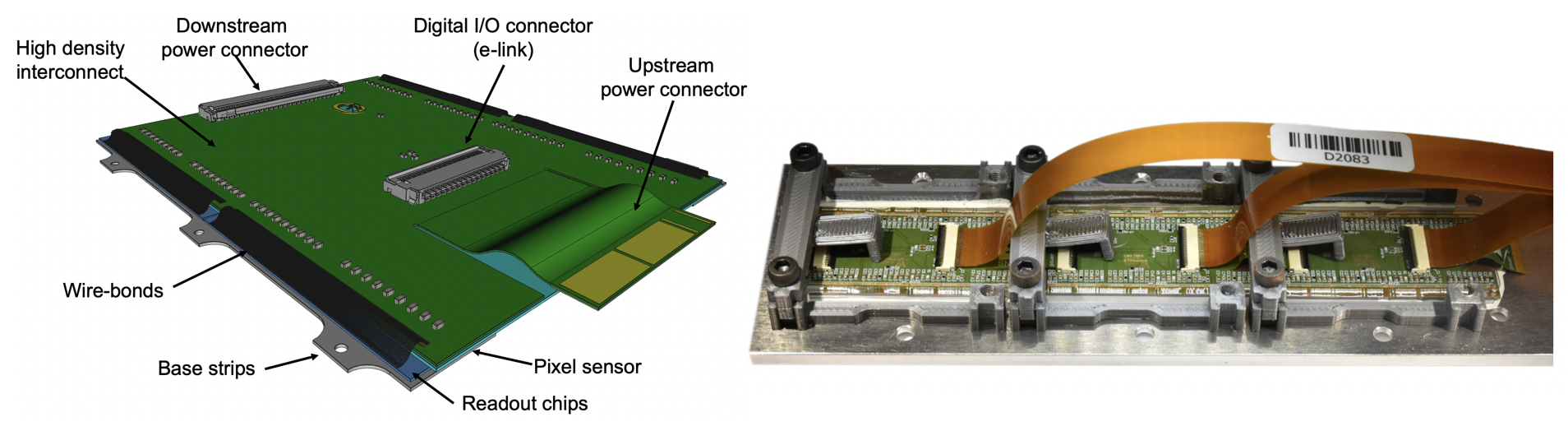}
\caption{Sketch of a 2\,$\times$\,2 pixel module (left), and a serially powered chain of three  digital modules (right).}
\label{fig:IT-module}
\end{figure}

The basic units of the IT detector are the hybrid pixel modules. As shown in Fig.\,\ref{fig:TK-layout}, two types of pixel modules will be employed in the detector: 1\,$\times$\,2  (green) and  2\,$\times$\,2 (yellow) readout chips (ROC) per sensor.  The module consists of ROCs bump-bonded onto the sensor that is glued on a High-Density-Interconnect flexible PCB (HDI). The connection between the ROCs and HDI is made via wire-bonds and the ROCs are mounted on aluminum nitride base strips  as Fig.\,\ref{fig:IT-module} shows. The HDI is designed to provide a return supply current path and for low material budget. It is equipped only with power and readout connectors and passive components used for setting some of the ROCs operational parameters.

An extensive silicon pixel sensors R\&D for qualifying sensors capable to operate up to the  $10^{16}$\,n$_{\mathrm{eq}}$\,cm$^{-2}$ range is currently on-going and 150\,$\mu$m thick n-in-p planar sensors are the current baseline for the IT detector. In order to guarantee a consistent charge collection efficiency after irradiation the technology requires a high bias voltage up to $\sim$1\,kV. Due to the sensor structure and the high voltage presence in the proximity to the ROC, a spark protection between the two components is required. A dedicated R\&D on the use of a combination of three types of dielectric protection (BenzoCycloButene layer on sensor edges, polyimide layer on ROC periphery and parylene coating at module level) started in 2019 and a reliability study after irradiation is currently on-going with the target to converge to a solution at the end of 2020. 
An alternative for the inner layers of the TBPX is the 3D silicon sensor technology, which is more radiation hard and requires lower bias voltages. The 3D technology has the disadvantage of having a larger pixel capacitance and a more complex fabrication process with a lower yield. 

Two possible pixel sizes are being considered: 25\,$\times$\,100\,$\mu$m$^{2}$, as current baseline, and 50\,$\times$\,50\,$\mu$m$^{2}$.  Initial studies show that, the relative difference  in term track parameter resolution between the two designs is rather small with a trade-off between primary vertex discrimination and resolution on the impact parameter. For the innermost layer, square pixels would result in a very long cluster  which would set more stringent requirements on the operational threshold of the chip, and it would also demand a larger bandwidth for reading the data. 

The readout chip for the ATLAS and CMS pixel detectors at HL-LHC is under design by the CERN RD53 collaboration~\cite{RD53}. The chip, designed in 65\,nm CMOS technology, has been developed with a cell size of 50\,$\times$\,50\,$\mu$m$^{2}$, a low threshold (below 1000\,e$^{-}$), a high data rate (one 160 Mb/s input link and up to four 1.28 Gb/s output links) and serial powering capability. 
The first half-size prototype chip, RD53A~\cite{RD53A}, featured a 400\,$\times$\,192 pixel matrix and has been used for extensive R\&D and qualification programs by ATLAS and CMS. Radiation hardness up to 500 Mrad was proven with indications that operation up to 1\,Grad would be possible under controlled conditions, most importantly by cooling the chip during the full lifetime. 
In the RD53A chip, three different analogue front-ends (FE) are prototyped in three sub-matrices. These are the synchronous FE, linear FE  and differential FE. After a dedicated review process CMS has selected the linear FE for the final chip.
The common design framework of the final chip, known as RD53B, contains design improvements and a few fixed bugs identified in the RD53A. A dedicated overview of the qualification results is given in~\cite{RD53B}.
The CMS final chip size is 16.8\,$\times$\,21.6\,mm$^{2}$ with a matrix of 336\,$\times$\,432 pixels. The submission of the first full size chip prototype for CMS is expected at the end of 2020.

\begin{figure}[h!]
\centering
\includegraphics[scale=0.47]{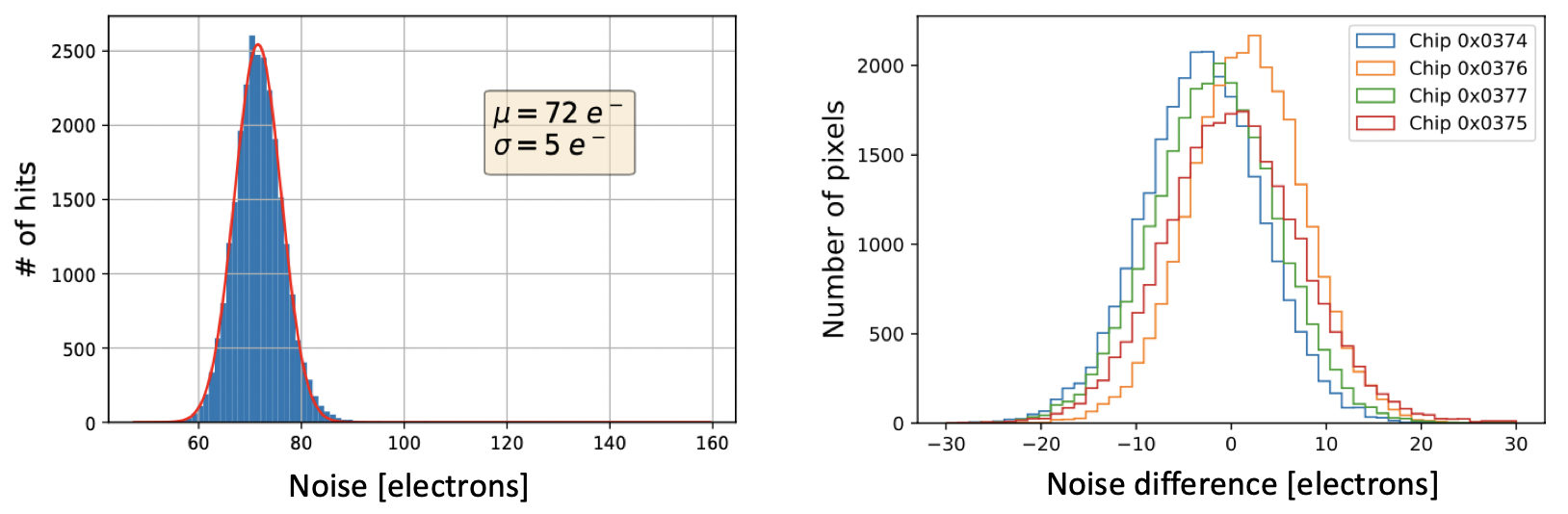}
\caption{Typical ENC distribution in electrons of an individually powered readout chip, e.g. RD53A, (left) and per-pixel difference after inclusion in a serially powered chain (right) \cite{SPplots}.}
\label{fig:SPplots}
\end{figure} 

The extreme rate requirements for the readout chip necessitate the use of a CMOS technology with low supply voltage ($\sim$1.2\,V), resulting in a chip that must be supplied with significant current levels ($\sim$2\,A per chip). It has been calculated that the total needed power for the Inner Tracker is around 50\,kW.
The use of a standard detector powering schema that has all readout chips supplied in parallel with a constant voltage is not possible, due to the voltage drop along the cables and the prohibitively large cable cross section that would be required. Similarly point-of-load-DCDC conversion cannot be employed due to space constraints.
The only viable powering scheme for such an environment is serial distribution. In this scheme 8 to 12 modules are arranged in a chain. All chain elements receive the same current and the voltage is equally shared if all elements represent the same and constant load. This is possible due to the ShuntLDO implementation in the readout chip  that combines a linear regulator (LDO) and a shunt. A picture of a serially powered chain of three digital modules is shown in Fig.\,\ref{fig:IT-module}, while Fig.\,\ref{fig:SPplots} shows the noise distribution of a module in standalone operation and the chip by chip difference after inclusion in the serial chain~\cite{SPplots}.

The IT modules will be  connected with up to 1.6\,m electrical links to optical-modules located at the periphery of the detector. The optical module hosts two LpGBT~\cite{LpGBT} transceivers and two VTRx+~\cite{VTRX} optical links. In total  six 1.28 Gb/s up-links per module will be implemented for data and monitoring information, and one 160 Mb/s down-link for bringing clock, trigger, fast commands and configuration data to the module.

%%%%%%%%%%%%%%%%%%%%%

\section{Outer Tracker system}

The Outer Tracker consists of six barrel layers and five endcap disks per end, and it is subdivided in: TB2S (Tracker Barrel with 2S modules), TBPS (Tracker Barrel with PS modules) and two TEDDs (Tracker Endcap Double Disks). 
The OT covers a surface of 190\,m$^{\mathrm{2}}$  with 13'200  modules featuring about 213 million of readout channels.

\begin{figure}[h!]
\centering
\includegraphics[scale=0.45]{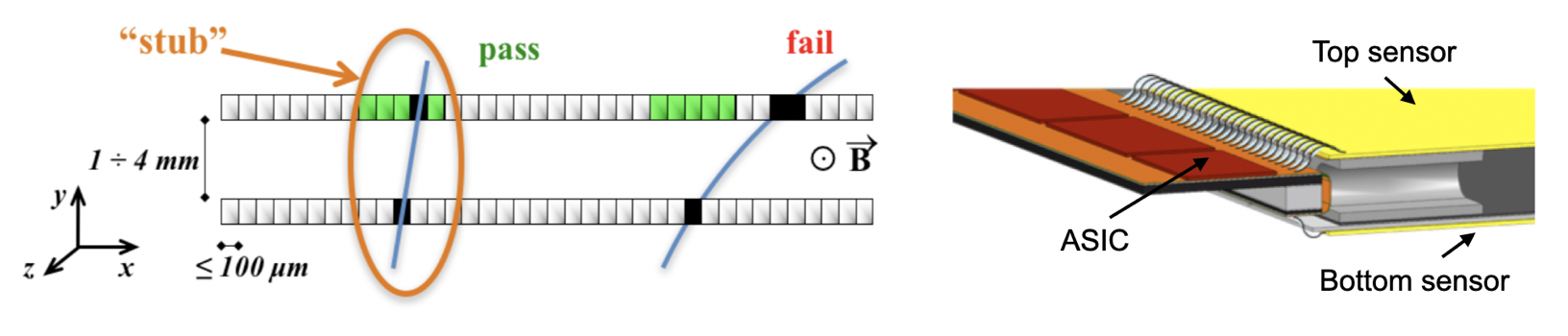}
\caption{Sketch of the track stub finding principle (left) and the  p$_{\mathrm{T}}$ module concept (right). A track passes both sensors of a module. A low momentum track falls outside the acceptance window and produces no stub~\cite{TKTDR}.}
\label{fig:pT-concept}
\end{figure}

The modules have the ability to autonomously select track segments (stubs) above a selected p$_{\mathrm{T}}$ threshold and to send these to the backend electronics. 
The selection relies on the bending of the charged particles in the magnetic field and a programmable selection window in the readout chips (see Fig.\,\ref{fig:pT-concept}).
The backend track finder system receives the stub data from the individual detector modules and performs track finding  in two steps, pattern recognition and track fitting, and sends the final tracks to the L1 trigger\,~\cite{TKTDR}.

\begin{figure}[h!]
\centering
\includegraphics[scale=0.6]{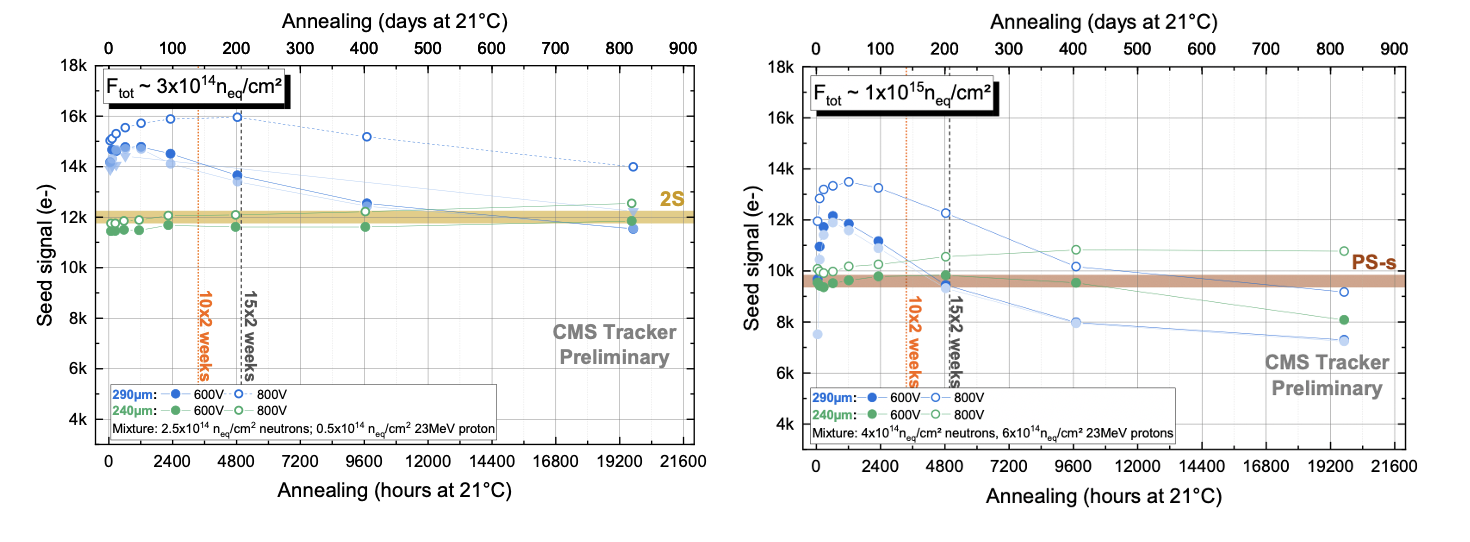}
\caption{Seed signal as a function of annealing time for FZ290 (blue) and thFZ240 (green) strip sensors after irradiation to the maximum expected fluences after 3000\,fb$^{-1}$ for the 2S (left) and PS (right) modules. Solid (open) points refer to signal collected with a bias voltage of 600 (800) V. The horizontal lines represent the required signal for the 2S and the PS strip (PS-s) sensors.}
\label{fig:SeedSignal}
\end{figure}

An extensive R\&D was undertaken to identify the sensor technology for the OT. At the end, only two options were left: Float Zone (FZ) silicon n-in-p with an active sensor thickness of 290\,$\mu$m (FZ290) or 240\,$\mu$m (FZ240). The sensor technology chosen for both sensors of the p$_{\mathrm{T}}$ modules is the FZ290.  An extensive irradiation and characterization program showed that FZ290 provides sufficient seed signal at the standard operation voltage of 600\,V and the expected maximum fluence after 3000\,fb$^{-1}$. For scenarios up to and beyond 4000\,fb$^{-1}$, an increase to an operation voltage of 800\,V would allow FZ290 to maintain adequate performance even at the most exposed locations. Beam test performance of strip sensors prototype are described in~\cite{Sensors1},~\cite{Sensors2}. 
An overview of the seed signal as a function of annealing time for the strip sensors after irradiation to the maximum expected fluences is shown in Fig.\,\ref{fig:SeedSignal}.

\begin{figure}[h!]
\centering
\includegraphics[scale=0.43]{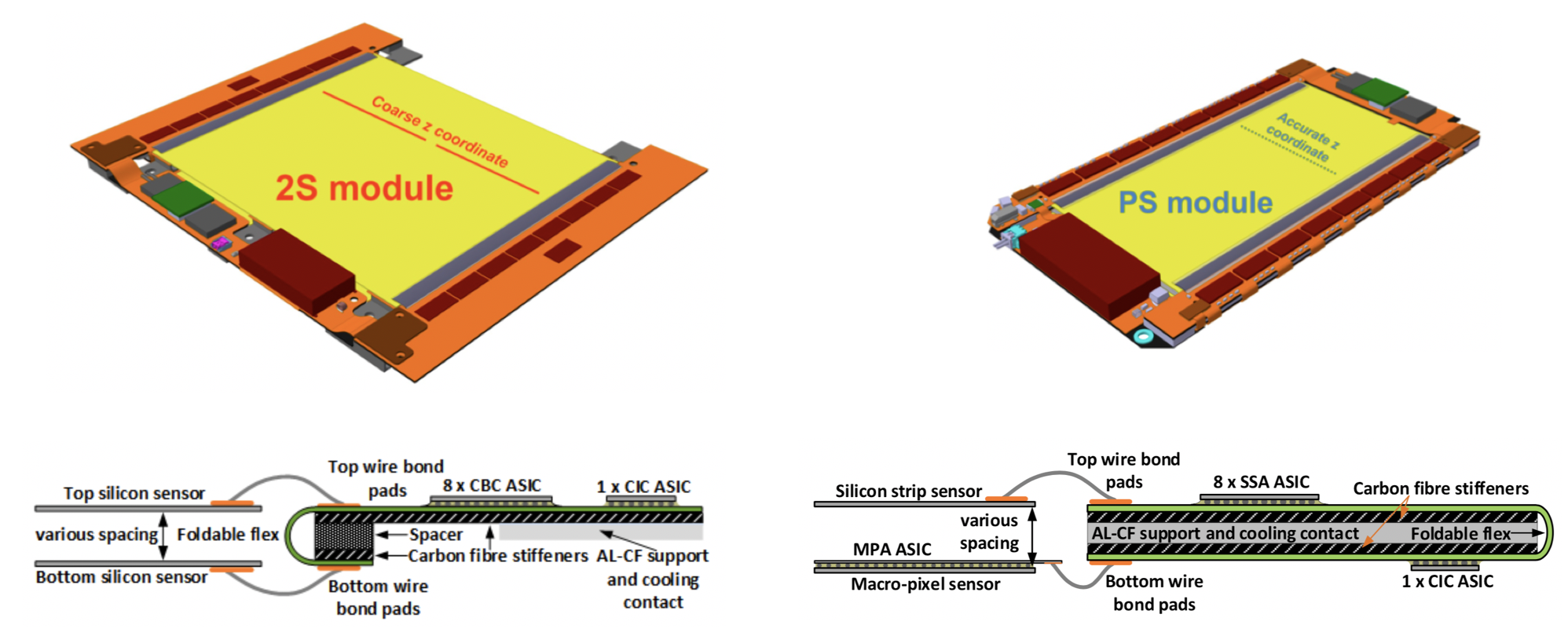}
\caption{The 2S module (left) and PS module (right) of the Outer Tracker. Shown are views of the assembled modules (top), and sketches of the front-end hybrid folded assembly and connectivity (bottom) ~\cite{TKTDR}.}
\label{fig:pT-modules}
\end{figure}

Both p$_{\mathrm{T}}$ modules have two different types of high density interconnect hybrid  circuits which house the front-end and auxiliary electronics. The hybrids are stand-alone units that are connected via bidirectional optical links to the backend electronics with no intermediary aggregator system.
The front-end hybrids host the readout chips and the concentrator chip while the service hybrids house the opto-electronics and the DC-DC converters for powering. Each side of the module is connected to a front-end hybrid. For both module types the signals from the top and bottom sensor are routed to one readout chip to perform the track stub finding. This is achieved by using a flexible hybrid which is folded over a spacer and which allows routing of the signal between the different parts of the module.
A sketch of both module types with their front-end and other hybrids is shown in the Fig.\,\ref{fig:pT-modules}.

The 2S module features two silicon strip sensors each with two columns of 1016 strips  with single strip size of 5\,cm\,$\times$\,90\,$\mu$m. Each sensor side is read out by eight  readout chips (CBC) implemented in 130\,nm CMOS technology. A CBC chip reads 254 strips (127 from bottom and 127 from top sensor strips), performs hit correlation 
between the two sensors and sends the stub data out at each bunch crossing to the concentrator chip (CIC, designed in 65\,nm CMOS technology), that performs data sparsification, formats the output data, and sends them to the service hybrid. The service hybrid hosts the LpGBT, VTRx+ optical link, DC-DC converters and HV distribution circuitry. The data from the front-end hybrids are merged and sent via a single optical fibre to the back-end electronic system.

The PS module is made of one silicon micro-strip sensor with two columns of 960 strips each with single strip size of 2.5\,cm\,$\times$\,100\,$\mu$m, and a macro-pixel sensor with a matrix of 32\,$\times$\,960 pixels with a pixel size of 1.5\,mm\,$\times$\,100\,$\mu$m. The strip sensor is read out by two times eight short strip ASIC (SSA), while the macro-pixel sensor is read out by sixteen macro pixel ASIC (MPA); both chips are implemented in 65\,nm CMOS technology. 
The track stub finding in the PS modules is done by the MPA. The MPA chip receives the information about strip clusters from the SSA which sends them together with information about the bunch crossing in which a hit occurred. The MPA combines this information with the macro-pixel information to form track stubs. The transfer scheme and data formats are very similar to those used in the 2S module, allowing the same CIC chip to be used for performing the same data collection, sparsification and formatting functions as for the 2S module. For space reasons, the PS module has two service hybrids, one for the optical system and one for the powering.

The module design uses novel composite materials (Al-CF) and all structural components have been chosen for their high thermal conductivity, low CTE, and minimal material budget. 

\begin{figure}[h!]
\centering
\includegraphics[scale=0.5]{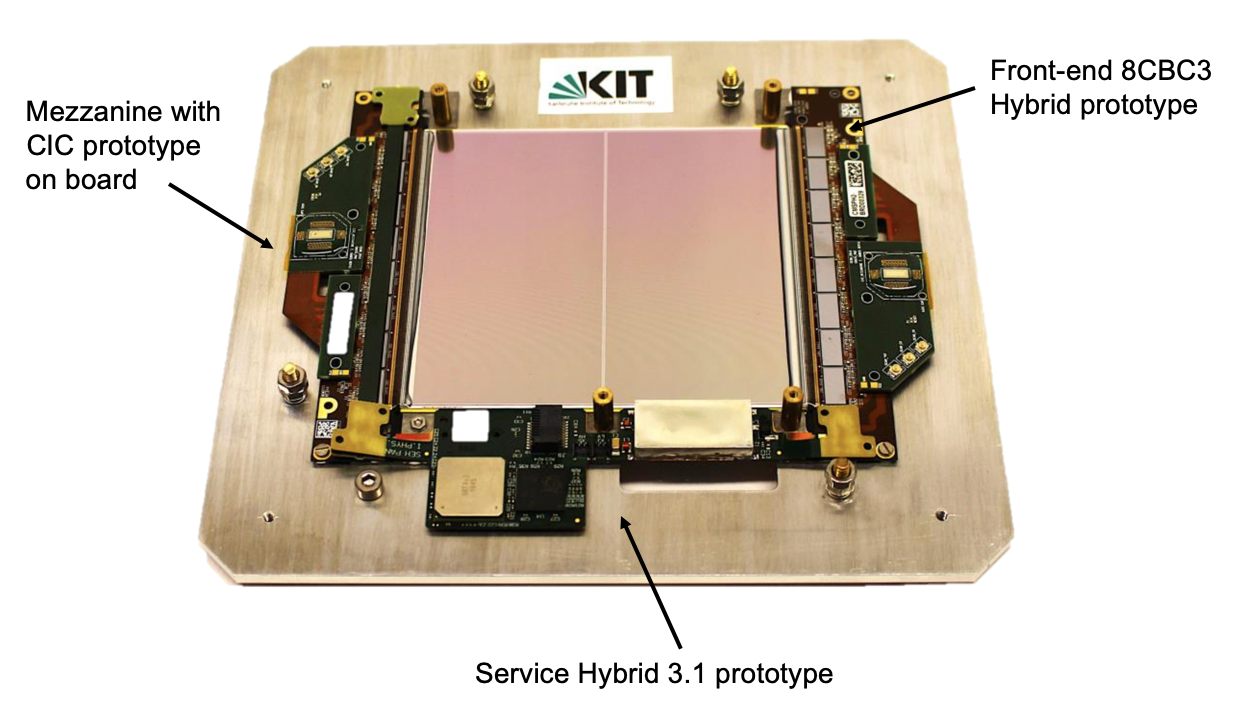}
\caption{Photo of a recent 2S module prototype.}
\label{fig:2Smodule}
\end{figure} 

The module prototyping (assembly and testing)  has been successfully accomplished in different CMS institutes and, as an example, Fig.\,\ref{fig:2Smodule} shows a recent 2S module built with the latest prototype components. Details on the performance of 2S modules in test-beam environment are described in ~\cite{2Stestbeam}.

%%%%%%%%%%%%%%%%%%%%%

\section{Material budget and performance}
\begin{figure}[h!]
\centering
\includegraphics[scale=0.28]{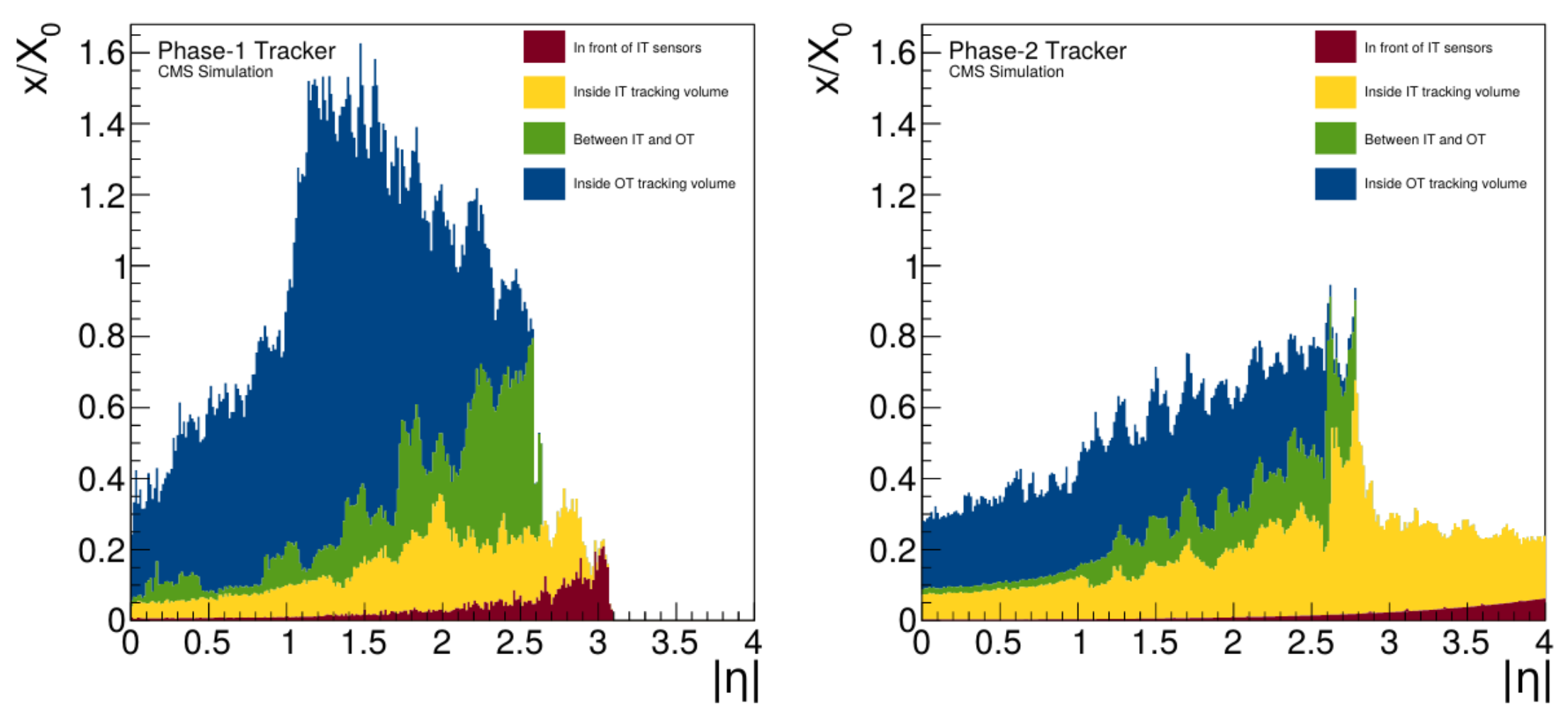}
\caption{Material budget for the currently installed CMS tracker (left) and the new tracker (right).~\cite{TKTDR}.}
\label{fig:material-budget}
\end{figure}

Despite the  increased number of readout channels in the new tracker the estimated material budget shows a significant reduction compared to the currently installed tracker as is shown in Fig.\,\ref{fig:material-budget}. The key features to achieve this reduction are: a reduced number of layers, an optimised routing of the services, use of light weight material, low-mass CO$_{2}$ cooling as well as the use of DC-DC converters (OT) and serial powering (IT).

\begin{figure}[h!]
\centering
\includegraphics[scale=0.26]{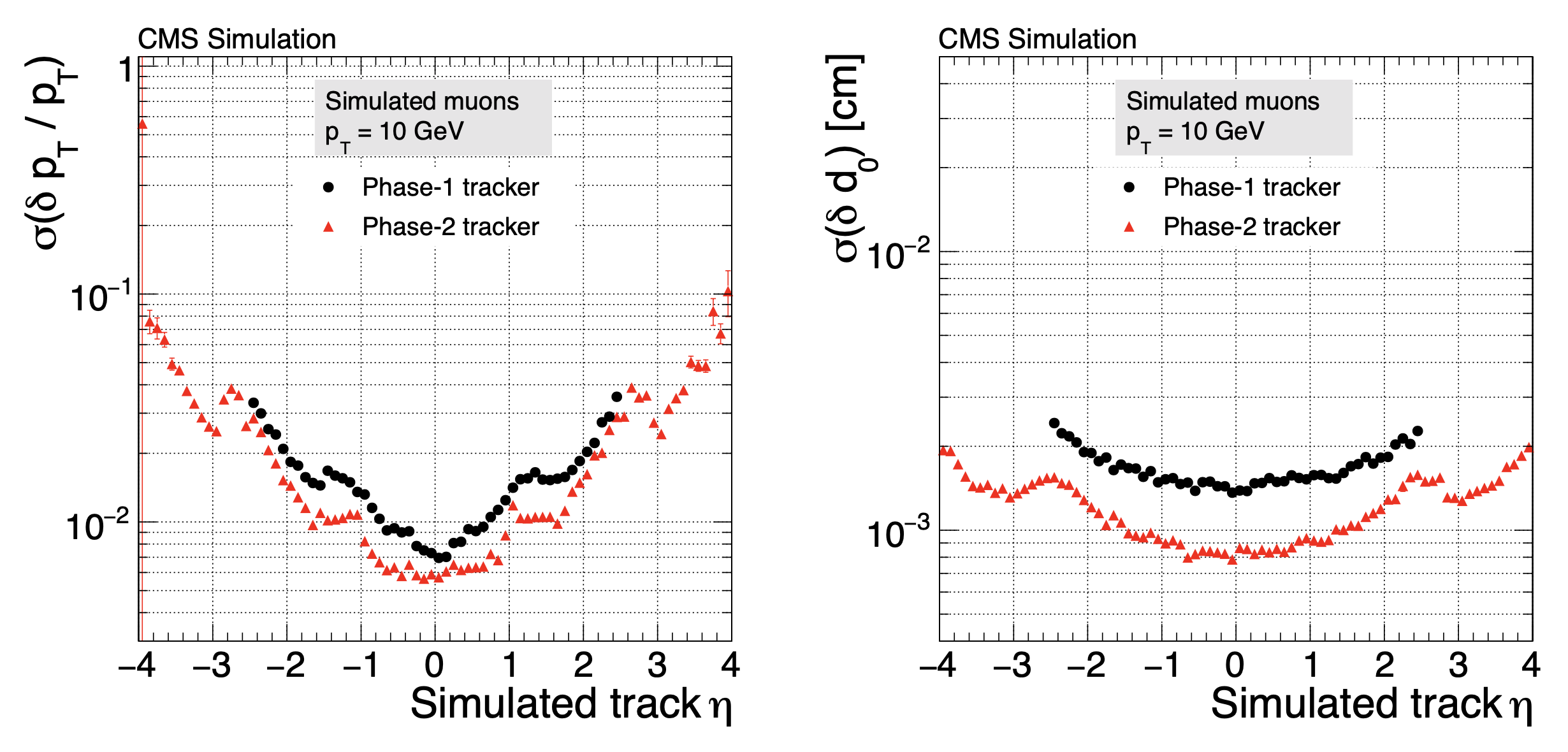}
\caption{Relative resolution of the transverse momentum (left) and resolution of the transverse impact parameter (right) as a function of the pseudorapidity for the Phase-1 (black dots) and the new  Phase-2 tracker (red triangles), using single isolated muons with a transverse momentum of 10\,GeV ~\cite{TKTDR}.}
\label{fig:performances-1}
\end{figure}

Also the tracking and vertexing capabilities of the new tracker will be better than for its predecessor.  Figure\,\ref{fig:performances-1} shows  the p$_{\mathrm{T}}$ resolution (left) and impact parameter resolution (right) comparison between the currently installed tracker (Phase-1) vs.\,the new tracker (Phase-2), respectively.

\begin{figure}[h!]
\centering
\includegraphics[scale=0.38]{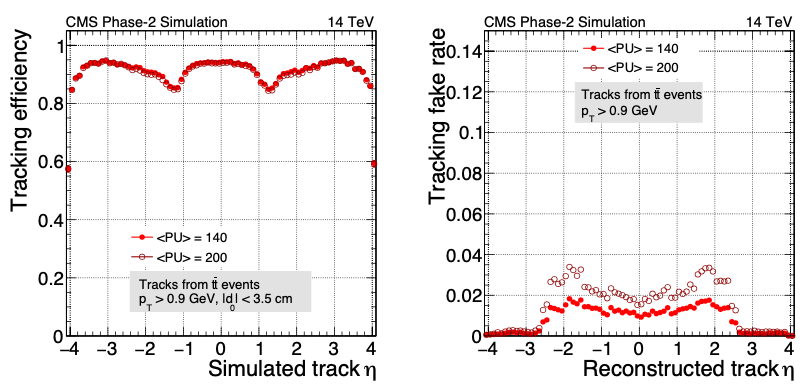}
\caption{ Tracking efficiency (left) and fake rate (right) as a function of the pseudorapidity for tt-bar events with 140 pileup events (full circles) and 200 pileup events (open circles). The tracks are required to have p$_{\mathrm{T}}$ higher than 0.9 GeV. The efficiency is shown for tracks produced within a radius of 3.5 cm from the centre of the luminous region.~\cite{TKTDR}.}
\label{fig:performances-2}
\end{figure}

As shown in Fig.\,\ref{fig:performances-2}, the new tracker is also expected to be capable of maintaining a high tracking efficiency  (about 90\%) at the high pile-up and have a fake rate below few percent.

%%%%%%%%%%%%%%%%%%%%%

\section{Summary}
The CMS Phase-2 Tracker is an ambitious project that has to cope with a higher pile-up and radiation environment at the HL-LHC. The new tracker will consist of about 200\,m$^{2}$ of silicon and will be made up of the Outer Tracker using modules containing pairs of closely spaced sensors, and the Inner Tracker with silicon pixel sensors. 
The key features of the new tracker are the high granularity, radiation hardness, low material budget and the capability to provide tracking information to the first stage of the CMS trigger system.
The project is overall in good shape and  on-track for installation in LS\,3 with the Inner Tracker entering the prototyping phase and the Outer Tracker in prototyping phase with first items entering the pre-production phase.

\end{document}